\documentclass[12pt,twoside,fleqn]{article}
\usepackage{epsfig}
\usepackage{multirow}
\usepackage{exscale}
\def\lsim{\raise0.3ex\hbox{$<$\kern-0.75em\raise-1.1ex\hbox{$\sim$}}}
\def\gsim{\raise0.3ex\hbox{$>$\kern-0.75em\raise-1.1ex\hbox{$\sim$}}}
\makeatletter
\@addtoreset{equation}{section}
\makeatother

\renewcommand{\thefootnote}{\alph{footnote}}

\newcommand{\beq}{\begin{equation}}
\newcommand{\eeq}{\end{equation}}
\newcommand{\bqa}{\begin{eqnarray}}
\newcommand{\eqa}{\end{eqnarray}}
\newcommand{\bqas}{\begin{eqnarray*}}
\newcommand{\eqas}{\end{eqnarray*}}
\newcommand{\bdm}{\begin{displaymath}}
\newcommand{\edm}{\end{displaymath}}
\newcommand{\Ns}{N_{\sigma}}
\newcommand{\Nt}{N_{\tau}}

\newcommand{\nn}{\nonumber}

\newcommand{\hmu}{\hat \mu}
\newcommand{\hnu}{\hat \nu}
\newcommand{\hro}{\hat \rho}
\newcommand{\hsi}{\hat \sigma}
\newcommand{\hs}{\hat s}
\newcommand{\hc}{\hat c}
\begin{document}

\thispagestyle{empty}
\renewcommand{\thefootnote}{\fnsymbol{footnote}}
 \mbox{} \hfill BNL-NT-03/3\\
 \mbox{} \hfill BI-TP 2003/07\\
 \mbox{} \hfill March 2003\\
%
\begin{center}
{{\large \bf Infinite Temperature Limit of Meson Spectral Functions \\[2mm]
Calculated on the Lattice}
 } \\

\vspace*{1.0cm}
{\large F. Karsch$^1$, E. Laermann$^1$, 
P. Petreczky$^2$\footnote[1]{Goldhaber fellow} and S. Stickan$^1$}

\vspace*{1.0cm}
{\normalsize
$\mbox{}$ {$^1$ Fakult\"at f\"ur Physik, Universit\"at Bielefeld,
D-33615 Bielefeld, Germany}\\
$\mbox{}$ {$^2$ Nuclear Theory Group, Physics Department
Brookhaven National Laboratory, Upton NY 11973}
}
\end{center}
\vspace*{1.0cm}
\centerline{\large ABSTRACT}

\baselineskip 20pt

\noindent
We analyze the cut-off dependence of mesonic spectral
functions calculated at finite temperature on Euclidean
lattices with finite temporal extent. In the infinite
temperature limit we present analytic results for lattice
spectral functions calculated with standard Wilson fermions
as well as a truncated perfect action. We explicitly determine
the influence of `Wilson doublers' on the high momentum
structure of the mesonic spectral functions and show that
this cut-off effect is strongly suppressed when using an
improved fermion action.

\vfill
\eject

\renewcommand{\thefootnote}{\alph{footnote}}
\setcounter{footnote}{0}

\baselineskip 15pt

\section{Introduction}

The thermal modification of basic properties of hadrons, e.g. their
masses and decay widths, is one of the central issues in the discussion
of experimental signals that can emerge from the dense partonic systems
created in heavy ion collisions. Lattice calculations can, in
principle, provide this information through the analysis of thermal
properties of Euclidean correlation functions of suitably chosen
operators carrying hadronic quantum numbers. These correlation functions
contain all the necessary information on the temperature dependence
of hadronic spectral functions. In order to make such
studies quantitative and reliably extract information valid in the
continuum limit we have to understand, however, in detail the cut-off
dependence of spectral functions calculated on the lattice at finite
temperature. We provide here a detailed analysis of mesonic
spectral functions in the infinite temperature limit and discuss
their cut-off dependence. This provides a basis for discussions
of the cut-off dependence of spectral functions at finite temperature
and is similar in spirit to studies of the cut-off dependence of the QCD
equation of state which have first been performed in the
ideal gas (infinite temperature) limit \cite{Kar82}.

Information on the hadronic spectrum is extracted in lattice
calculations from properties of Euclidean time correlation functions
of suitably chosen hadronic currents. It has been suggested recently
\cite{Nak99} to apply the Maximum Entropy Method (MEM), a well known
statistical tool for the analysis of noisy data \cite{Bryan}, also
to the analysis of these correlation functions.
This opened the possibility to extract detailed information on
hadronic spectral functions, $\sigma (\omega,T)$, at zero as well as finite
temperature
\cite{Asa00,Wet00}. The first studies of spectral functions
based on the MEM approach
[6-13]
have indeed been encouraging.
These first studies, however, also showed that it is necessary to
get control over typical lattice problems like finite cut-off
effects or the influence of fermion doublers on the energy
dependence ($\omega$) of spectral functions. 
Cut-off effects show up in the large energy regime of spectral functions
and one generally may not be too much worried about them when one is
interested in extracting information about the low energy part of the spectral
functions. This, however, is different for the analysis of properties of
heavy quark
bound states and also is of particular importance for studies performed at
finite temperature. In the latter case the high energy part of, e.g.
the vector spectral function, is directly related to physically
observable dilepton cross sections and has been studied in much
detail in (resummed) perturbative calculations \cite{Braaten}. Moreover,
in the plasma phase of QCD
quasi-particle excitations are heavy and typically will have masses
which rise proportional to the temperature. At high temperature this
is expected to lead to broad resonance-like structures in spectral functions.
In lattice calculations, which at present all have been performed
with Wilson type fermion formulations, it is in general difficult
to distinguish such effects from contributions arising
from so-called heavy Wilson doublers. On a more technical level it also is important
for the MEM analysis to include information on the short distance behaviour
of correlation functions in the default model. This requires
information on lattice cut-off effects in the spectral function as
well as possible modifications of the integration kernel.

There are apparently plenty of reasons to get control over the
lattice cut-off effects in numerical calculations of spectral
functions. We will analyze these in the infinite temperature limit
of QCD by explicitly calculating hadronic spectral functions on
lattices with finite temporal extent $N_\tau$. We will present
results for spectral functions calculated on isotropic as well as 
anisotropic lattices and discuss their quark mass dependence.
Moreover, we will present results for standard Wilson fermions as 
well as a truncated perfect action \cite{biet1,biet2}.

The paper is organized as follows. In Section 2 we summarize known
results for free quark-antiquark spectral functions calculated in the
continuum. In Section 3 we perform the corresponding calculation
for Wilson fermions on the lattice. Section 4 is devoted
to a discussion of these lattice spectral functions, in particular
their quark mass and anisotropy dependence. In Section 5 we
present results from a calculation with an improved Wilson fermion
action, a truncated perfect action. Finally we give our conclusions in
Section 6. Some details of our calculations are presented in two Appendices.

\section{Thermal quark-antiquark spectral functions}

Thermal quark-antiquark correlation functions in coordinate space,
$G_H(\tau,\vec{x})$, are defined as
\begin{eqnarray}
G_H(\tau,\vec{x}) &=&
\langle J_H (\tau, \vec{x}) J_H^{\dagger} (0, \vec{0}) \rangle \quad ,
\label{eq:mesoncor}
\end{eqnarray}
where $\langle ... \rangle$ denotes the thermal average. The local sources 
for currents with different mesonic quantum numbers $H$ are given by
$J_H (\tau,\vec{x}) =\bar{q}(\tau, \vec{x})\Gamma_H q(\tau, \vec{x})$,
and $\Gamma_H$ is an appropriate combination of $\gamma$-matrices;
{\it i.e.,} $\Gamma_H = 1$, $\gamma_5$, $\gamma_\mu$, $\gamma_\mu \gamma_5$
for scalar,
pseudo-scalar, vector and pseudo-vector channels, respectively.
From these we obtain the mixed correlation functions at fixed
momentum $\vec{p}$ which are commonly considered in lattice calculations,
\begin{equation}
G_H(\tau, \vec p) = \int d^3 \vec x \; G_H(\tau, \vec x)\; 
e^{i \vec x \vec p} \quad .
\label{mixedcor}
\end{equation}
These two point functions have a spectral representation,
\begin{equation}
G_H(\tau, \vec p) = \int_{0}^{\infty} d \omega\;
\sigma_H(\omega, \vec p,T) \; K(\omega,\tau) \quad ,
\label{specrep}
\end{equation}
where $\sigma_H(\omega, \vec p,T)$ denotes the temperature dependent spectral
function and $K(\omega,\tau)$ is the integration kernel which carries
the entire dependence on Euclidean time $\tau\in [0,1/T)$,
\beq
K(\omega,\tau) =
\frac{\cosh[\,\omega(\tau-1/2T)]}{\sinh(\omega/2T)} \quad .
\label{kernel}
\eeq
It is easy to convince oneself that the spectral function appearing in
Eq.~\ref{specrep} indeed is the Minkowski space spectral function (see e.g.
\cite{LeBellac})
\begin{eqnarray}
\sigma_H(\omega, \vec p,T) =
{1 \over Z(T)} \sum_{n,m}&&\hspace*{-0.9cm}
e^{-E_n(\vec p)/T} (1-e^{-\omega/T})\;
\delta(\omega+E_n(\vec p)-E_m(\vec p))\cdot \nonumber\\
&&
\hspace*{-0.9cm}{|\langle n | J_H(0)| m\rangle|}^2,
\label{specrepdelta}
\end{eqnarray}
where $Z(T)$ is the partition function and $\langle ... \rangle$ denotes here
the matrix element of the hadronic current $J_H(0)$ taken between
energy eigenstates at fixed momentum $\vec{p}$.

The correlation functions can be evaluated using the momentum space
representation of the quark propagator and its spectral representation
\cite{LeBellac}. In the following we will mainly be concerned with the
zero momentum spectral functions,
$\sigma_H(\omega,T)\equiv\sigma_H(\omega,\vec p=\vec 0,T)$, which take on
a rather simple form \cite{LeBellac,Mustafa},
\begin{eqnarray}
\sigma_H(\omega,T) = \frac{N_c}{8 \pi^2} \,
\Theta(\omega^2- 4 m ^2)\, \omega^2 \, \tanh\left(\frac{\omega}{4T}\right)
\sqrt{1-\left(\frac{2m}{\omega}\right)^2}\; \cdot
\nonumber \\
\cdot\; \left[a_H + \left(\frac{2m}{\omega}\right)^2 b_H \right] 
+ \frac{N_c}{3} \frac{T^2}{2} f_H\;
\omega \; \delta (\omega)
\quad .
\label{eq:freecontinuum}
\end{eqnarray}
Here $N_c$ denotes the number of colours, eg. $N_c=3$.
For some selected mesonic quantum number channels, $H$,
the coefficients $a_H,~b_H$ and $f_H$ are given in Table~\ref{tab:coeff}.
We note that in some cases a contribution proportional to a $\delta$-function
at vanishing energy appears in Eq.~\ref{eq:freecontinuum}. At non-zero
temperature this gives rise to a constant, $\tau$-independent term in the
Euclidean correlation function defined in Eq.~\ref{specrep}.
For massless quarks, also at non-vanishing momentum,
$p = \sqrt{\vec p^{\,2}}$, a rather compact form for the spectral function is obtained,
\begin{eqnarray}
\sigma_H(\omega,\vec p,T) = \frac{N_c}{8 \pi^2} \, (\omega^2- p^{\,2})
\, a_H \,
\left\{
\Theta(\omega^2- p^{\,2}) \,
\frac{2T}{p}
\ln\frac{\cosh(\frac{\omega+p}{4 T})}
        {\cosh(\frac{\omega-p}{4 T})} \right.\nonumber \\
 +
\left.
\Theta( p^{\,2} - \omega^2)
\left[ \frac{2T}{p}
\ln\frac{\cosh(\frac{p+\omega}{4 T})}
        {\cosh(\frac{p-\omega}{4 T})}
-\frac{\omega}{p} \right]
\right\} \quad .
\end{eqnarray}

\begin{table}[t]
\begin{center}
{\begin{tabular}{|c|c|c|c|c|c|}
\hline
\rule[-2mm]{0mm}{5mm}$H$ & $a_H$ & $b_H$ & $f_H$ & $c_H^{\rm lat}$ & $d_H^{\rm lat}$ \\
\hline
\rule[-2mm]{0mm}{5mm}$PS$                    & 1 & 0 & 0 & 1   & 0    \\
\rule[-2mm]{0mm}{4mm}$S$                     &-1 & 1 & 0 & -d  & d-1  \\
\hline
\rule[-2mm]{0mm}{5mm}$V_0$                   & 0 & 0 & -1 &  0   & -1   \\
\rule[-2mm]{0mm}{6mm}$\displaystyle{\sum_{i=1}^3 V_i}$      & 2  & 1  & $1$ &3-d & d    \\
\rule[-2mm]{0mm}{5mm}
$\displaystyle{V\equiv \sum_{\mu=0}^3 V_\mu}$  & 2 & 1 & 0 & 3-d & d-1  \\
\hline
\rule[-2mm]{0mm}{5mm}$A_0$                  & 0  & 0  & 1 & 1-d & d    \\
\rule[-2mm]{0mm}{6mm}$\displaystyle{\sum_{i=1}^3} A_i$ & -2   & 3  & -1  & -2d &
2d-3 \\
\rule[-2mm]{0mm}{5mm}
$\displaystyle{A\equiv \sum_{\mu=0}^3} A_\mu$ &-2 & 3 & 0 & 1-3d & 3(d-1) \\
\hline
\end{tabular}}
\end{center}
\caption{Coefficients $a_H$, $b_H$ and $f_H$ for free spectral functions
in different mesonic quantum number channels $H$ (Eq.~\ref{eq:freecontinuum}).
The last two columns give coefficients appearing in the definition of the
corresponding correlation functions on the lattice (Eq.~\ref{corrfkt}).
The momentum dependent function $d$ is defined in Eq.~\ref{eq:d}.
}
\label{tab:coeff}
\end{table}

Due to asymptotic freedom of QCD the free field limit
is approached at infinite temperature.
In order to discuss the infinite temperature or free field limit of spectral
functions
and correlation functions it is appropriate to rescale all 
variables with non-trivial dimension by appropriate powers of the 
temperature, e.g.
$\tilde{\omega}=\omega/T$. For fixed $\tilde{m}\equiv m/T$ the rescaled
correlation functions,
$\tilde{G}_H (\tilde{\tau},\tilde{\vec p}) \equiv 
G_H(\tau T, \vec p/T) / T^3$,
then have a well defined infinite temperature limit,
\begin{equation}
\tilde{G}_H (\tilde{\tau},\tilde{p})
= \int_{0}^{\infty} d \tilde{\omega}\;
\tilde{\sigma}_H(\tilde{\omega}, \tilde{\vec p},T) \;
\tilde{K}(\tilde{\omega},\tilde{\tau}) \quad ,
\label{rescaledG}
\end{equation}
where $\tilde{\sigma}$ denotes the rescaled spectral density,
$\tilde{\sigma}_H(\tilde{\omega},\tilde{\vec p},T) =
\sigma_H (\omega,\vec p,T)/T^2$, and $\tilde{K}(\tilde{\omega},\tilde{\tau})
=\cosh(\tilde{\omega}(\tilde{\tau}-1/2))/\sinh(\tilde{\omega}/2)$.

Eventually we are interested in obtaining information
on the temperature dependence of the physical spectrum and we thus want
to determine the Minkowski space spectral function as introduced
in Eq.~\ref{specrepdelta}. 
In the continuum the spectral function is connected to the
temporal Euclidean correlation function via Eq.~\ref{specrep}.
The analyticity properties required to make that connection,
however, in general only exist in the continuum.
Therefore, extracting a spectral function from lattice
data on Euclidean correlation functions via Eq.~\ref{specrep}
will suffer from
lattice artifacts and it is only in the continuum limit that a direct
relation to the spectral properties at finite temperature can be
established. 
This will become clear in the following section where
we discuss the spectral representation of mesonic correlation functions
using the standard Wilson fermion formulation on Euclidean lattices with
finite temporal extent $N_\tau$.

\section{\hspace{-0.2cm}Lattice Spectral Functions with Wilson Fermions}

In the following,
the dimensionless representation
of hadronic two point correlation functions and the corresponding rescaled
spectral functions given in Eq.~\ref{rescaledG}
will be analyzed for their cut-off dependence.
The situation here is similar to the discussion
of the cut-off dependence of bulk thermodynamic variables, e.g. the
rescaled pressure, $P/T^4$ \cite{Kar82,pressure2}.
In that case, on isotropic lattices, deviations from the continuum
result can be expressed in terms of the lattice spacing given in units of
the temperature, $aT$, which is nothing else but the inverse of the
temporal extent of the lattice, $aT = 1/N_\tau$. 
For bulk thermodynamic observables the temperature is in general the
only\footnote{This is correct in the limit of vanishing as well as infinite
quark masses.}
quantity with non-trivial dimension that can set the scale for the cut-off dependence.
In the case of spectral functions, however, the energy  provides
another scale and we can expect to find an additional dependence of
the spectral functions on $a\omega = \tilde{\omega}/N_\tau$. However,
we will show in the following explicitly that only $\tilde{\omega}/N_\tau$
determines the cut-off dependence and a sole dependence on
$1/N_\tau$ does not appear in the spectral functions.
On anisotropic lattices cut-off effects are, in addition, controlled by
the ratio $\xi=a/a_\tau$ of spatial ($a$)
and temporal ($a_\tau$) lattice spacings.
In this case, temperature and energy in (spatial) lattice units 
also depend on this ratio as 
$a T = \xi/N_\tau$ and $a \omega = \tilde \omega \xi / N_\tau$,
respectively.

We will discuss here the cut-off dependence of spectral functions
calculated 
within
a generalization of Wilson's fermion
discretization scheme \cite{Wilson} on anisotropic lattices. In the
free field limit the fermion action is diagonal in the colour degrees
of freedom, $S_F = \sum_{k,l}\sum_{c=1}^3 \bar{\psi}_k^c\; Q_{k,l}\;
\psi_l^c$ with the fermion matrix
\begin{eqnarray}
Q_{k,l} = &&\hspace*{-1.0cm}
\biggl[ r_\tau +  \frac{3r + \hat m}{\xi}  \biggr]
\delta_{k,l}
\;-\; \frac{1}{2}\biggl[ (r_\tau-\gamma_0)\delta_{k+\hat{0},l}
+ (r_\tau+\gamma_0)\delta_{k-\hat{0},l}\biggr]
\nonumber \\
&-&\frac{1}{2\; \xi}\sum_{i=1}^{3}\biggl[ (r-\gamma_i)\delta_{k+\hat{i},l}
+ (r+\gamma_i)\delta_{k-\hat{i},l}\biggr] \quad .
\label{Waction}
\end{eqnarray}
where the dimensionless quark mass, $\hat m$, is expressed
in units of the spatial lattice spacing,
$\hat m = m a$.
The generic choice of the Wilson action is $r=r_\tau=1$ \cite{Wilson}.
We will discuss here
only the case $r_\tau =1$ as this avoids the occurrence of a
second pole in the fermion propagator which would give rise to an unphysical
time-like doubler mass.
We will, however, consider the space-like Wilson
$r$-parameter, $r\in (0,1]$,
as an additional free parameter
of the fermion action. 
Even on anisotropic lattices with $\xi\; >\; 1$
the generic choice for the $r$-parameter is $r=1$.
Yet, by choosing $r=1/\xi$
discretization errors of order $m a$
can be completely eliminated at leading order from
meson correlators with improved Wilson (clover) fermions
on anisotropic lattices \cite{rchoice},
e.g. in studies of heavy quark systems.
As this
choice has, in fact, been used in recent studies of heavy quark
spectral functions \cite{Umeda02} we also will explore the dependence
of spectral functions on the choice of $r$.

Analytic results for
the free field limit of hadronic correlation functions on isotropic
lattices of size $N_\sigma^3 \times N_\tau$ have been presented in
the past using free Wilson fermions \cite{Carpenter} with $r=1$. It is
straightforward to extend these calculations to the case of anisotropic
lattices and $r\ne 1$. Starting point
for the calculation of hadronic correlation functions is the
momentum space representation of the free Wilson fermion
propagator,
\begin{equation}
S(k) =
  \frac{-i\gamma_0\sin k_0-i{\cal K}+[(1-\cos k_0)+{\cal M}]}
  {\sin^2k_0+{\cal K}^2+[(1-\cos k_0)+{\cal M}]^2} \quad ,
\label{quarkprop}
\end{equation}
with
\begin{eqnarray}
  {\cal K}&=&\frac{1}{\xi} \sum \limits_{i=1}^3 \gamma_i \sin k_i \quad , \\
  {\cal M}&=&\frac{1}{\xi} \biggl[r \sum \limits_{i=1}^3 ( 1-\cos k_i ) \;+\;
\hat m \biggr] \quad .
\end{eqnarray}
On a finite lattice of size
$N_\sigma^3\times N_\tau$ the momenta take on discrete values,
$k_0 = \frac{2\pi}{N_\tau} (n_0+1/2)$ with $n_0=0,~\pm 1,..\pm (N_\tau/2
-1)$, $N_\tau /2$ and $k_i = \frac{2\pi}{N_\sigma} n_i$ with
$n_i=0,~\pm 1,..\pm (N_\sigma /2-1)$, $N_\sigma /2$ for  $i=1,~2,~3$.

Following Ref.~22
one finds for the temporal zero momentum free quark-antiquark
correlation functions with mesonic quantum numbers, $H$,

\begin{equation}
\hspace*{-0.4cm}\tilde{G}_H(\tilde{\tau},\tilde{p}\equiv0) =
N_c \left( {\frac{N_\tau}{\xi N_\sigma}} \right)^3
  \sum \limits_{{\vec k}} \;
  \frac {c_H^{lat}({\vec k}) \cosh[2E({\vec k}) N_\tau(\tilde{\tau}-1/2)]+
d_H^{lat} ({\vec k})} {(1+{\cal M})^2 \cosh^2(E({\vec k})N_\tau/2)}
~,
\label{corrfkt}
\end{equation}
which is defined on the discrete set of Euclidean times
$\tilde \tau  = n/ N_\tau$, with $n = 0,~1,..., N_\tau-1$, 
accessible on a lattice with temporal extent $N_\tau$.
The energy, 
$E\equiv E({\vec k})$, is given by the location of
the pole of the denominator of the Wilson fermion propagator 
(Eq.~\ref{quarkprop}) at $ik_0 = E({\vec k})$, {\it i.e.}
\begin{equation}
\cosh E\; =\; 1+\frac{{\cal K}^2+{\cal M}^2}{2(1+{\cal M})}  \quad .
\label{energy}
\end{equation}
The functions $c_H^{lat}$ and $d_H^{lat}$ appearing in Eq~\ref{corrfkt}
depend on the three momentum ${\vec k}$ through the function
\begin{equation}
d\equiv d({\vec k}) =  {{\cal K}^2 \over \sinh^2 E} \quad .
\label{eq:d}
\end{equation}
Note that $d$ is approaching 1 in the continuum limit.
For some quantum number channels explicit expressions are given in
Table~\ref{tab:coeff}.

Eqs.~\ref{corrfkt}-\ref{eq:d} can be used to analyze the
infinite temperature limit of mesonic correlation functions on
any finite lattice of size $N_\sigma^3 \times N_\tau$.
In the following we will take the thermodynamic limit\footnote{In
general we found that the dependence on the spatial
extent becomes weak for $\xi N_\sigma /N_\tau \gsim 3$.} ($N_\sigma
\rightarrow \infty$) and concentrate on the cut-off
dependence of these correlation functions which arises from $N_\tau$
being finite. In the thermodynamic
limit the momenta are continuously distributed in the interval
$[-\pi,\pi]$ and the energy consequently becomes a continuous
function. This allows for
an integral representation of Euclidean
correlation functions in complete analogy to the continuum
representation given in Eq.~\ref{specrep}. In particular, we will
show that also on lattices with finite temporal extent $N_\tau$
the integration kernel is identical to the continuum kernel,
Eq.~\ref{kernel}. Cut-off effects which
are responsible for the deviation of the lattice correlation
functions from the corresponding continuum correlation functions
thus only show up in the lattice spectral functions.

In the thermodynamic limit the momentum sum appearing in Eq.~\ref{corrfkt}
gets replaced by a three-dimensional integral over the lattice
Brillouin zone, \\[2mm]
\centerline{$\displaystyle{\frac{1}{N_\sigma^{3}} \sum_{\vec k}\quad
\rightarrow \quad
\int \limits_{{\vec k}} \equiv \frac{1}{(2\pi)^{3}} \int\hspace*{-0.3cm}
\int \limits_{-\pi}^{\pi} \hspace*{-0.3cm}\int  {\rm d}^3{\vec k}}$ \quad .}
\\[2mm]
It is obvious from Eq.~\ref{corrfkt} and Table~\ref{tab:coeff} that there
will appear only two types of $\tau$-dependent integrals which result from
the two terms appearing in the definition of $c_H^{lat}$, {\it i.e.} 1 and
$d({\vec k})$ respectively,
\begin{eqnarray}
\tilde{G}_1(\tilde{\tau}) &=&N_c\;
  \left( {\frac{N_\tau}{\xi}} \right)^3
  \int \limits_{{\vec k}} \frac {1}{(1+{\cal M})^2}
  \frac{\cosh[2E({\vec k})N_\tau(\tilde{\tau} -1/2)]}{\cosh^2(E({\vec k})
  N_\tau/2)} \quad , \nonumber \\
\tilde{G}_2(\tilde{\tau}) &=&N_c\;
  \left( {\frac{N_\tau}{\xi}} \right)^3
  \int \limits_{{\vec k}} \frac {d({\vec k})}{(1+{\cal M})^2}
  \frac{\cosh[2E({\vec k})N_\tau(\tilde{\tau} -1/2)]}{\cosh^2(E({\vec k})N_\tau/2)}
\quad .
\label{G1G2}
\end{eqnarray}
In addition there are also two $\tau$-independent integrals
related to the sums involving the term $d_H^{lat}$ appearing in
Eq.~\ref{corrfkt}.
These constants contribute to the $\delta$-functions at vanishing
frequency.
On finite lattices all quantum number channels
listed in Table~\ref{tab:coeff} (except the pseudoscalar)
will receive non-vanishing
contributions $f_H^{lat}$,

\begin{equation}
f_H^{lat} = N_c \biggl( \frac{N_{\tau}}{\xi } \biggr)^3
\int_{\vec k} \frac{d_H^{lat}(\vec k)}{(1+{\cal M})^2 }
\frac{1}{\cosh^2(E({\vec k})N_{\tau}/2)} \quad .
\label{fHlat}
\end{equation}

The integrals given in Eq.~\ref{G1G2} are the starting point for
our discussion of the cut-off dependence of mesonic spectral functions.
So far we have achieved to represent the correlation functions
$\tilde{G}_H(\tilde{\tau})$ in terms of three-dimensional
momentum integrals. Our goal is to find, for finite $N_\tau$, a
spectral representation defined through the one-dimensional
integral given in Eq.~\ref{specrep}.
This can be achieved by introducing 
the energy in units of the temperature,
$\tilde{\omega} = 2 EN_\tau$,
as one of the integration variables.
In Appendix A we show explicitly for the case $r=1$ the sequence of
variable transformations required to obtain an integral
representation which is in complete analogy
to the continuum relation, Eq.~\ref{rescaledG}, {\it i.e.} we can
write the integrand of this integral as a product of a
$\tau$-independent {\it lattice spectral function} expressed in
units of $T^2$ and a $\tau$-dependent kernel which is identical with
the continuum kernel $\tilde{K}$ defined in Eq.~\ref{kernel},
\begin{eqnarray}
  \label{integraleq}
\tilde{G}_i(\tilde{\tau})=
\int \limits_{\tilde{\omega}_{min}}^{\tilde{\omega}_{max}}
  {\rm d}\tilde{\omega} \; \tilde{\sigma}_i(\tilde{\omega},N_\tau)\;
\tilde{K}(\tilde{\omega}, \tilde{\tau} )
\quad , \quad i\; =\; 1,\; 2 \quad .
\end{eqnarray}
The spectral functions, $\tilde{\sigma}_i(\tilde{\omega},N_\tau)$, explicitly
depend on the lattice cut-off, which is reflected in the explicit dependence
on $\tilde \omega / N_\tau = \omega a_\tau$,

\begin{eqnarray}
  \tilde{\sigma}_1(\tilde{\omega},N_\tau) &=&
  \frac{N_c}{2\pi^3}\frac{N_\tau^2}{\xi^3}\;
  \tanh\left( { \frac{\tilde{\omega}}{4} }\right) \;
  \sinh\left( {\frac{\tilde{\omega}}{4N_\tau} }\right)
  \sinh\left( {\frac{\tilde{\omega}}{2N_\tau} }\right)
  \;I_1(\tilde{\omega}/N_\tau,\xi) \nonumber \\
  \tilde{\sigma}_2(\tilde{\omega},N_\tau) &=&
  \frac{N_c}{2\pi^3}\frac{N_\tau^2}{\xi^3}\;
  \tanh\left( { \frac{\tilde{\omega}}{4} }\right) \;
  \frac{\sinh^3\left( {\frac{\tilde{\omega}}{4N_\tau} }\right) }
  {\sinh\left( {\frac{\tilde{\omega}}{2N_\tau} }\right) }
  \;I_2(\tilde{\omega}/N_\tau,\xi) \quad .
\label{latspec}
\end{eqnarray}
The two-dimensional integrals $I_i(\tilde{\omega}/N_\tau,\xi)$ are
worked out in Appendix A for the case $r=1$. They can,
however, also be defined for arbitrary values of $r$.

The integration limits in Eq.~\ref{integraleq} depend on the quark mass,
the anisotropy
and the Wilson $r$-parameter. For $r=1$ the maximal energy is determined
by the largest quark three momentum possible, ${\vec k}=(\pi,\pi,\pi)$, and
we find from Eq.~\ref{energy},
\begin{equation}
\tilde{\omega}_{min} = 2N_\tau\ln \left( {1+\hat m/\xi }\right)
\quad , \quad
\tilde{\omega}_{max} = 2N_\tau\ln \left( {1+(6+\hat m)/\xi }\right) \quad .
\label{bounds}
\end{equation}
For $r < 1$, however, the maximal energy generally corresponds to a
momentum in the interior of the first Brillouin zone. The corners
of the 3-dimensional Brillouin zone are local minima of the dispersion
relation which are interpreted as doubler masses which are proportional
to the Wilson $r$-parameter, {\it i.e.} they become lighter with
decreasing $r$. As we will see this leads to rather
complicated spectral properties even in the free quark, infinite temperature
limit.

We also note that the lattice spectral functions $\tilde{\sigma}_i$ are
directly proportional to the massless spectral functions in the continuum
($\tilde{\sigma}_{H} \sim \tilde{\omega}^2 \tanh (\tilde{\omega}/4)$).
In the massless limit the deviations from the continuum results thus
only arise through the ratio $\tilde{\omega}/N_\tau\equiv \omega
a_\tau$ which is the energy expressed in units of the temporal lattice
spacing. Eq.~\ref{latspec}
explicitly reflects a well known feature of the lattice formulation,
{\it i.e.} cut-off effects depend on the energy scale.
Of course, an explicit dependence on the lattice spacings
$(a,a_{\tau})$ never appears in the lattice formulation.
It is, however, remarkable that also no explicit dependence on
$1/N_\tau \equiv a_\tau T$ appears in the spectral functions.

Finally, we want to reconstruct from $\tilde{\sigma_i}$ the spectral
functions in fixed quantum number channels. In particular, we will
consider spectral functions in the pseudo-scalar, scalar, vector and
axial-vector channels which are given by
\begin{equation}
  \tilde{\sigma}_{PS}^{lat} = \tilde{\sigma}_1 \quad , \quad
  \tilde{\sigma}_{S}^{lat} = -\tilde{\sigma}_2 \quad , \quad
  \tilde{\sigma}_{V}^{lat} = 3\tilde{\sigma}_1-\tilde{\sigma}_2 \quad , \quad
 \tilde{\sigma}_{A}^{lat} = \tilde{\sigma}_1 - 3\tilde{\sigma}_2 \quad .
\end{equation}
We ignore here a term proportional to $\tilde{\omega} \delta (\tilde{\omega})$
which, as discussed above, arises from the $\tau$-independent part 
($f_H^{lat}$) of the correlation functions. For the above
quantum number channels the coefficients of the $\delta$-functions will
vanish in the continuum limit and we have checked that they are indeed
small already on lattices with temporal extent $N_\tau \sim 10$.

\section{Cut-off effects on isotropic and anisotropic lattices}

We will analyze here in detail the lattice spectral functions derived
in the previous section for some choices of paramters ($r,~\xi,~m$) 
which have been used in recent studies of meson spectral functions at finite 
temperature.

\subsection{Massless Quarks on isotropic lattices: $\xi=1$, $m=0$}

\begin{figure}[htb]
  \begin{center}
\hspace{-0.5cm}
    \includegraphics[width=7.5cm]{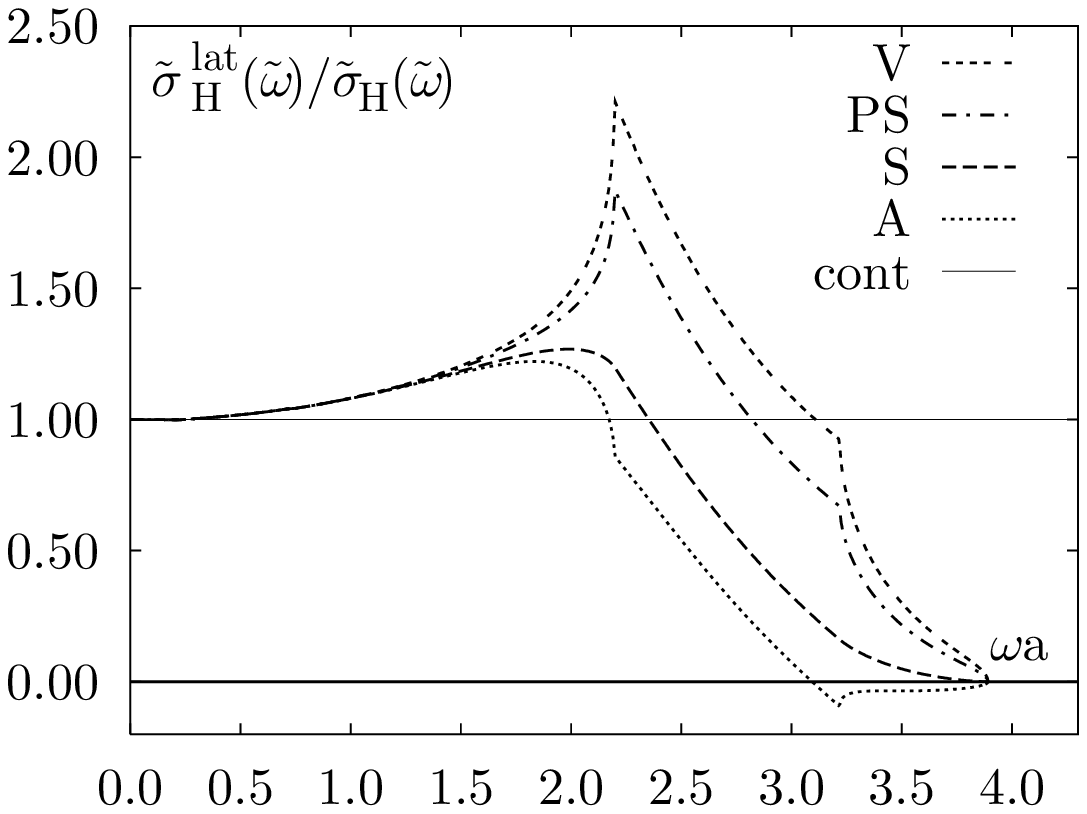}
    \includegraphics[width=7.5cm]{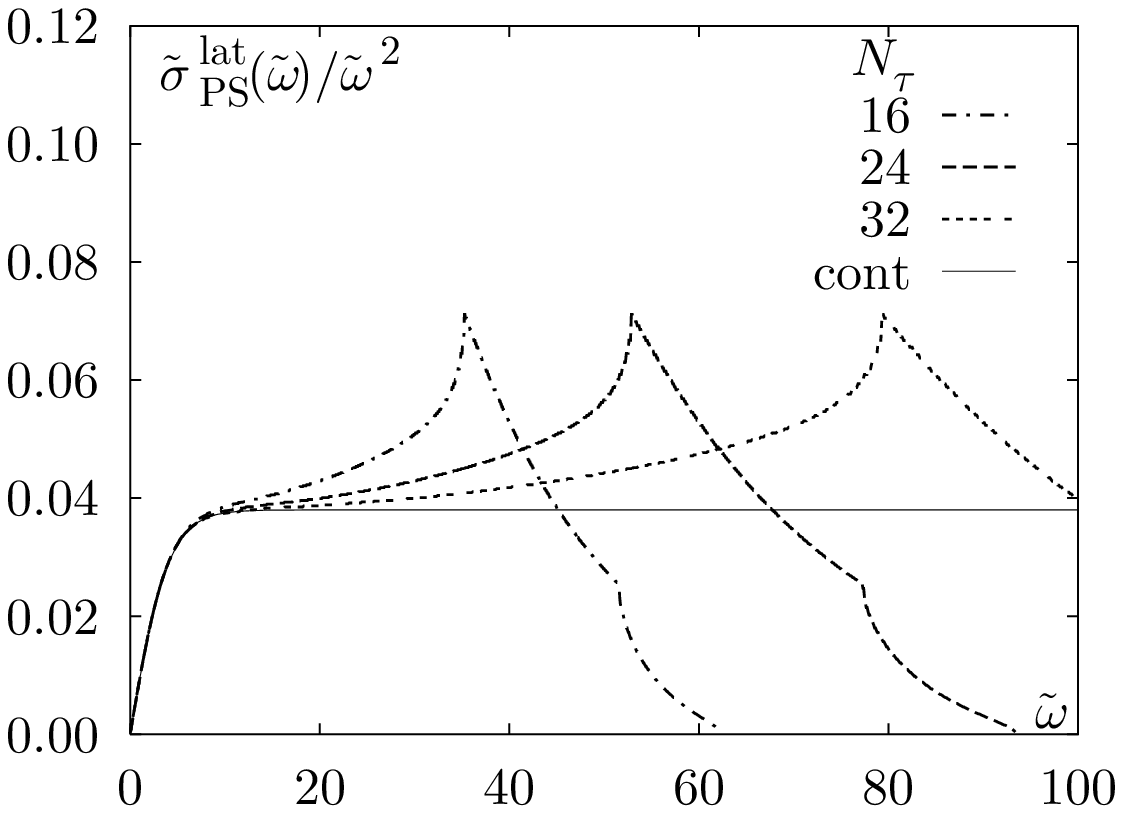}
    \caption{Ratio of lattice and continuum spectral functions
with $m=0$ and $\xi=1$ and for different quantum number channels (left).
The figure on the right hand side shows the pseudo-scalar lattice spectral
functions calculated  for lattices with temporal extent $N_\tau=16$, 24
and 32.  The solid line corresponds to the
continuum spectral function given in Eq.~\ref{eq:freecontinuum}.
    \label{fig:fSPF}
}
  \end{center}
\end{figure}

Let us first discuss the lattice size dependence of the spectral
functions for the case of vanishing quark masses ($m=0$) and on
isotropic lattices ($\xi = 1$). In Fig.~\ref{fig:fSPF}(left) we show
the ratios of the lattice and the corresponding
continuum spectral functions. We note that this ratio is a function
of $\tilde{\omega}/N_\tau  \equiv \omega a$ only.
The spectral functions vanish for $\tilde{\omega}
\ge \tilde{\omega}_{\rm max}$,
where $\tilde{\omega}_{\rm max} /N_\tau\equiv \omega_{max}a = 2 \ln 7$
is obtained from Eq.~\ref{energy}
as the largest energy possible for a mesonic state constructed from
two independent massless Wilson fermions with momentum
${\vec k}= (\pi,\pi,\pi)$. In the pseudo-scalar and vector
channels we observe a pronounced peak which occurs when the
momenta of both Wilson fermions correspond to the first
corner of the Brillouin zone ${\vec k}= (\pi,0,0)$. The corresponding
energy is $\omega_1a = 2\ln 3$. Finally we observe a cusp at
$\omega_2a = 2\ln 5$ between
$\omega_1a$ and $\omega_{\rm max}a$ which corresponds to the second
corner of the Brillouin zone ${\vec k}= (\pi,\pi,0)$. In the
continuum limit the lattice artifacts shift to higher energies and
the lattice spectral functions approach the continuum result
(Fig.~\ref{fig:fSPF}(right)).\footnote{
Note that in the scalar and axial case we always plot
the positive functions $-\sigma_S$ and $-\sigma_A$.}

Also in the interacting case \cite{CPpacs02} a peak-like structure has
been observed at similar values of the energy, {\it i.e.} for
$\omega a \simeq 1.7$ in the pseudo-scalar and  $\omega a \simeq 2$ in
the vector spectral function, respectively. It has been shown that these
peaks shift to larger energies when decreasing the lattice spacing. They 
thus have been identified as lattice artifacts which do not correspond to 
physical states in the continuum limit \cite{CPpacs02}.
It is likely that these peaks are remnants of the cut-off effects seen
here in the free spectral functions. To establish this relation in more detail 
it would certainly be interesting to analyze also 
spectral functions in the scalar and axial-vector channels.
At least in the free case the peaks which show up in the pseudo-scalar and
vector spectral functions 
are absent
in the scalar and axial-vector channels.

In Ref.~6
it has been suggested that the cut-off dependent
peaks in the spectral functions are related to bound states involving
heavy quark doublers with masses of ${\cal O}(1/a)$. 
In view of the free spectral function we would, however,
prefer not to speak of states at all. Rather,
the distortion
of the spectral function and the characteristic structures seen in
the free case are due to the 
lattice dispersion relation and to the
sudden restriction of available momentum
space that occurs when one of the fermion momenta reaches one of the
corners of the first Brillouin zone.

We also note that all four spectral functions coincide up to
$\omega a \simeq 1.5$ where they differ by less than 15\% from the
continuum result. This agreement of different quantum number channels
is reminiscent of the chiral symmetry of the free fermion action.
In the lattice formulation with Wilson fermions chiral symmetry is,
of course, explicitly broken  which leads to different
spectral functions in the scalar and pseudo-scalar sector.
As can be seen from Fig.~\ref{fig:fSPF} this explicit breaking most
strongly influences the large energy part of the spectrum, {\it i.e.}
$\sigma_{PS}$
deviates strongly from $ -\sigma_{S}$ for $\omega a\; \gsim\; 1.5$.
The same holds true in the vector and axial-vector channels. Also note that
for large energies the finite cut-off effects can lead to a negative
lattice spectral function in the axial-vector channel.

\subsection{Massless Quarks on anisotropic lattices: $\xi\; > \; 1$, $m=0$}

When reducing the temporal relative to the spatial lattice
spacing  $(\xi = a / a_\tau \; > \; 1)$ one has to increase the
number of grid points in the temporal direction if one wants to keep the
temperature constant, $1/T = N_\tau a_\tau$. In the interacting case
it requires a fine tuning of spatial and temporal couplings (hopping
parameter) in order to maintain rotational symmetry at zero temperature
and, of course, it will also increase the computational effort.
Nonetheless, it may be of advantage in
the analysis of mesonic correlation functions at high temperature
because one can make use of information on the correlation functions
at a larger number of Euclidean time steps.

\vspace{0.5cm}
\noindent
\underline{$r=1$}: \\[2mm]

\begin{figure}[h]
  \begin{center}
\hspace{-0.5cm}
    \includegraphics[width=9.5cm]{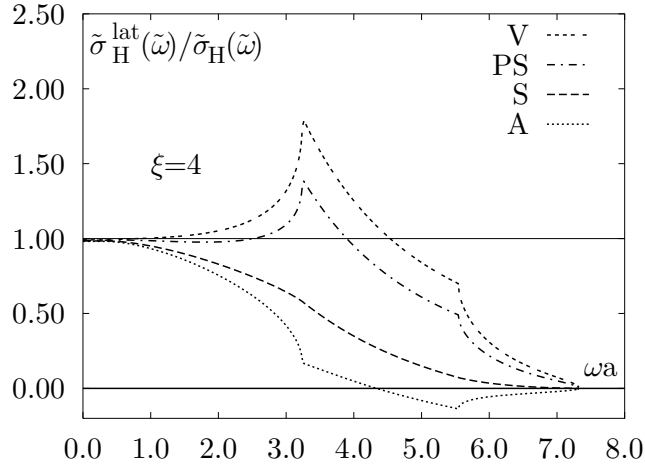}
    \caption{Ratio of lattice and continuum spectral functions for
    $m=0$ and $\xi=4$.
    \label{fSPFaniso}
}
  \end{center}
\end{figure}

Let us first discuss
the structure of free spectral functions on anisotropic lattices
for the case $r=1$.
As can be seen from Eq.~\ref{bounds} for a fixed ratio $\xi/N_\tau
\equiv Ta$ the support for the spectral
function increases with increasing anisotropy.
For $(N_\tau,\xi)\rightarrow \infty$  it reaches a finite limit,
{\it i.e.} $\tilde{\omega}_{min} \rightarrow 2\tilde{m}$
and $\tilde{\omega}_{max} \rightarrow 12 N_\tau / \xi + 2\tilde{m} $.
By increasing the temporal lattice size and the anisotropy
simultaneously the upper limit, above which the spectral functions vanish,
can thus be increased by about a factor 3 relative to the case of isotropic
lattices. For a moderate anisotropy factor ($\xi \simeq 4$), typically used
in numerical calculations, $\tilde{\omega}_{max}$ 
(and thus also $\omega_{max} a$ )
is about a factor 2
larger than in the isotropic case. Lattice artifacts, however, set in
earlier; for $\xi =4$ the peak in the pseudo-scalar and vector
spectral function is only shifted by a factor 1.4 (see Fig.~\ref{fSPFaniso}).

In Fig.~\ref{fSPFaniso} we show the same spectral functions as in 
Fig.~\ref{fig:fSPF} now calculated with an anisotropy $\xi =4$;
a choice of the anisotropy parameter which has been used in recent studies 
of spectral functions \cite{Asak02,Umeda02}. 
As can be seen from this figure the energy interval in which the
pseudo-scalar and vector spectral functions are only little affected
by cut-off effects is about twice as large as on the isotropic lattices.
Scalar and pseudo-scalar correlation functions, however, are affected 
differently.
As a consequence the degeneracy of both spectral functions, a precursor
of chiral symmetry restoration, is lifted at smaller energies than on
isotropic lattices. The situation is similar for the
vector and axial-vector spectral functions.

\vspace{0.5cm}
\noindent
\underline{$r=1/\xi$}: \\[2mm]

When one uses the particular choice
$r=1/\xi$ \cite{Umeda02} the doubler masses become light for large
$\xi$ and can influence the spectral properties at lower energies
than it is the case for $r=1$. As the quark dispersion relation
no longer leads to maxima in the corner of the Brillouin zone
the energy range in which the spectral functions are
non-zero shrinks compared to the $r=1$ case. For $\xi=1/r =4$ one finds
from the quark dispersion relation $\omega_{max}a =3.45$, which is
even slightly smaller than the corresponding value for $r=\xi =1$.

\begin{figure}[t]
  \begin{center}
\hspace{-0.5cm}
    \includegraphics[width=9.5cm]{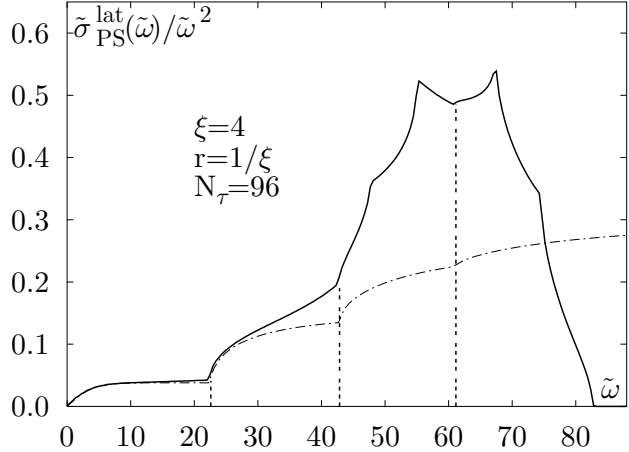}
    \caption{Pseudo-scalar lattice spectral function for massless quarks
calculated on a lattice with temporal extent $N_\tau = 96$
on an anisotropic lattice with $\xi=4$ and
Wilson $r$-parameter r$=1/\xi$. The dashed curve shows the
sum of contributions from
continuum spectral functions for one massless pseudo-scalar
and a number of massive pseudo-scalars constructed from free quarks with
masses corresponding to the doubler masses obtained from the quark 
dispersion relation of this action. The
horizontal lines indicate the location of doubler masses in the three
distinct corners of the first Brillouin zone.
    \label{fSPFani2}
}
\end{center}
\end{figure}

The non-monotonic behaviour of the dispersion relation also makes
it more complicated to find a closed analytic representation for the
spectral functions in terms of two-dimensional integrals as we have
done for the case $r=1$.
Although in principle it is possible to generalize the approach
described in Appendix A for $r=1$, 
in the case $r<1$ we have used the simpler numerical 
binning-approach to calculate 
spectral functions. This is also introduced in Appendix A.
The resulting pseudo-scalar spectral function is shown in
Fig.~\ref{fSPFani2} for massless quarks
and $\xi = 1/r = 4$. In this case the mass of the lightest, three-fold
degenerate doubler is $m_1 a = 0.471$ 
which gives rise to the first
threshold at $\tilde{\omega}_1 = 2 N_\tau m_1a /\xi$ seen in this figure. The
other structures seen in this figure result from contributions of
the doublers in the other corners of the Brillouin zone as well as
the maxima of the quark dispersion relation which now do not reside
in the corners of the Brillouin zone.

We note that this action does reproduce the continuum spectral
function well up to the point where the first doubler starts
contributing, $\omega a \simeq 1$. The energy range in which a
good agreement with the continuum spectral function can be
achieved thus is compatible with the isotropic case.

\subsection{Massive Quarks}

The modification of heavy quark bound states and, in particular,
their dissolution in a quark gluon plasma is considered to be an
important signature for the formation of dense matter in heavy ion
collisions. It thus is of interest to analyze also spectral functions
for heavy quark bound states in lattice calculations. The first
attempts to do so followed the strategies discussed in the
previous sections, {\it i.e.} calculations have been performed 
with Wilson or clover fermions on
isotropic \cite{Datta02} and anisotropic \cite{Umeda02} lattices.
In the latter case the formulation with $r=1/\xi$ has been used.

\begin{figure}[h]
  \begin{center}
\hspace{-0.5cm}
    \includegraphics[width=10.0cm]{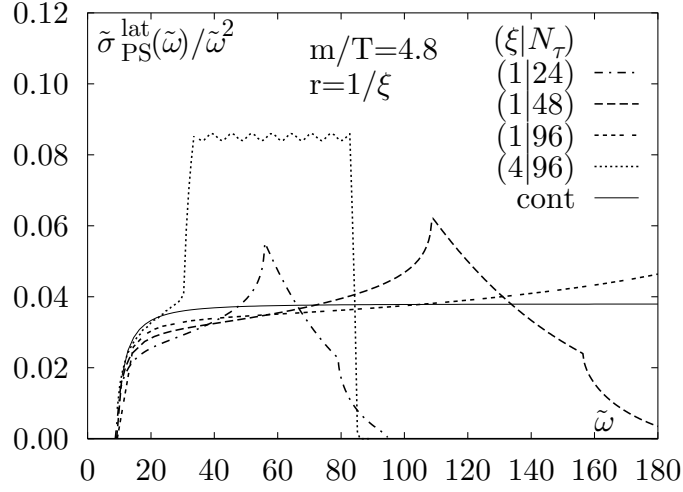}
    \caption{Free pseudo-scalar spectral functions for fixed $m/T$
calculated on isotropic lattices with different temporal extent
and quark masses. Also shown is the result obtained on an anisotropic
lattice with $\xi=1/r=4$. In this case the spectral function deviates
strongly from the continuum result for $\tilde{\omega}\gsim 30$ (see
also Fig.\ref{fSPFani2}). For better visibility we have cut-out this
part here and replaced it by a wavy line.
    \label{mfSPF}
}
  \end{center}
\end{figure}

The analysis presented in the previous sections for massless quarks
carries over also to the case of massive quarks. The structure of the
cut-off dependence discussed for the various types of actions
follows patterns similar to those seen in the massless case.
In the large energy region the cut-off effects dominate and a
non-vanishing quark mass has little influence on the location of the
pronounced peaks observed for $r=1$ or the additional thresholds
arising for $r=1/\xi$. This is shown in Fig.~\ref{mfSPF}
where we present results for quark masses $m/T=4.8$. Aside from the
low energy threshold which now is shifted to $\tilde{\omega}\simeq 2m/T$ 
the spectral
functions are similar to those shown in figures~\ref{fig:fSPF} and
\ref{fSPFani2}.

\section{Lattice Spectral Functions with improved Wilson Fermions}

\setcounter{figure}{0}
\setcounter{table}{0}

In the previous sections we have seen that spectral functions
obtained from calculations with the standard Wilson action reproduce 
the continuum
spectral functions at low energies, $\omega a \lsim 1.5$. Deviations
from the continuum dispersion relation, however, lead to strong
modifications of the spectral functions at larger energies. Moreover, the
violation of chiral symmetry becomes visible in spectral functions at these
energies. This motivates to analyze 
the cut-off effects of hadron correlation functions computed with
improved fermion actions which have better
chiral properties and lead to an improved dispersion relation. Much progress
has been made in constructing such actions
\cite{FPfermions}.
As a first step in this direction we want to analyze here a simple
truncated version of
a fixed point action \cite{biet1,biet2}. This action is constructed from a
small set of couplings which connect sites in an elementary
hypercube of the lattice and can be handled in close analogy to the
case of the Wilson action.
We restrict ourselves to the discussion
of the massless case on isotropic lattices. The action can be written as
\beq
S=\sum_{x,y} \bar \psi(x) \biggl\{ \sum_{\mu=0}^3\gamma_{\mu} \rho_{\mu}(x-y)
+\lambda(x-y)\biggr\} \psi(y) \quad ,
\label{perfect_s}
\eeq
with
\begin{eqnarray}
\rho_{\mu}(x-y)&=&\rho_1 (\delta_{y,x+\hmu}-\delta_{y,x-\hmu})+
 \sum_{\hnu \ne \hmu}   \rho_2 (\delta_{y,x+\hmu+\hnu}-\delta_{y,x-\hmu+\hnu})
\nonumber \\
&&+ \sum_{\hnu \ne \hmu \atop \hro \ne \hmu, \hnu}
  \rho_3 (\delta_{y,x+\hmu+\hnu+\hro}-\delta_{y,x-\hmu+\hnu+\hro})
\nonumber \\
&&+  \sum_{\hnu \ne \hmu \atop \hro \ne \hnu, \hsi \ne \hro}
  \rho_4 (\delta_{y,x+\hmu+\hnu+\hro+\hsi}-\delta_{y,x-\hmu+\hnu+\hro+\hsi})
\quad ,
\label{rho}
\end{eqnarray}
\begin{eqnarray}
\lambda(x-y)&=& \lambda_0 \delta_{y,x}+
\sum_{\mu} \lambda_1 (\delta_{y,x+\hmu}+\delta_{y,x-\hmu})+
 \sum_{\hnu \ne \hmu}
 \lambda_2 (\delta_{y,x+\hmu+\hnu}+\delta_{y,x-\hmu+\hnu})
\nonumber \\
&&+ \sum_{\hnu \ne \hmu  \atop \hro \ne \hmu, \hnu}
  \lambda_3 (\delta_{y,x+\hmu+\hnu+\hro}+\delta_{y,x-\hmu+\hnu+\hro})
\nonumber \\
&&+ \sum_{\hnu \ne \hmu \atop  \hro  \ne \hnu, \hsi \ne \hro}
  \lambda_4 (\delta_{y,x+\hmu+\hnu+\hro+\hsi}
      +\delta_{y,x-\hmu+\hnu+\hro+\hsi})
\quad .
\label{lambda}
\end{eqnarray}
Here $\hmu,\hnu,\hro$ and $\hsi$ denote unit vectors along positive 
as well as, except for $\hmu$,
negative directions in the hypercubic lattice.
Numerical values for the set of nine couplings
$\{ \rho_i \}_{i=1}^4$,
$\{ \lambda_i \}_{i=0}^4$ are given in Table 1 of \cite{biet2} for $m=0$.

Taking the Fourier transform of the action, Eq.~(\ref{perfect_s}), it is
straightforward to write down the propagator in momentum space
\bqa
\hspace{-0.8cm}S(k)=\frac{-i \gamma_0 \delta \sin k_0-i 
{\cal K}_1 -i {\cal K}_2\cos k_0+
            \kappa_1+\kappa_2 \cos k_0}{ ({\cal K}_1^2+ \kappa_1^2+\delta^2)
            +2 \cos k_0 ({\cal K}_1 {\cal K}_2+\kappa_1 \kappa_2)+
             \cos^2 k_0 ({\cal K}_2^2+\kappa_2^2-\delta^2)}~,
\label{prop_k}
\eqa
with
\bqa
  {\cal K}_1 = \sum_{i=1}^3 \gamma_i \alpha_i \quad , \quad
  {\cal K}_2 = \sum_{i=1}^3 \gamma_i \beta_i \quad .
\eqa
Explicit expressions for the momentum dependent
functions $\alpha_i=\alpha_i(\vec k)$,
$\beta_i=\beta_i(\vec k), \delta=\delta(\vec k)$, $\kappa_1=\kappa_1(\vec k)$
and $\kappa_2=\kappa_2(\vec k)$ are given in Appendix B.

\begin{figure}[t]
\begin{center}
\includegraphics[width=3.5in]{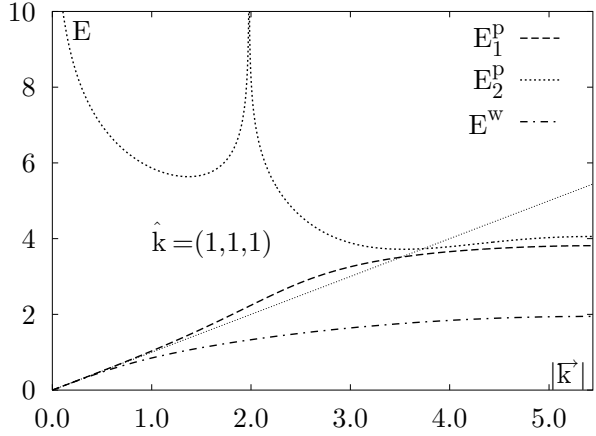}
\end{center}
\vspace*{-5mm}
\caption{The physical ($E_1^p$) and unphysical ($E_2^p)$ poles of the quark
propagator obtained from the truncated perfect action as the function of
the spatial momentum $\vec k$. The straight line shows the continuum
dispersion relation and the dash-dotted curve gives the dispersion
relation for the standard Wilson fermion action ($E^w$).
\label{perfect_disp}
}
\end{figure}

For the analysis of meson correlation functions and their spectral
representation
one has to calculate the quark propagator in the mixed $(\tau,\vec k)$
representation. The calculational steps are completely analogous to those
for the standard Wilson action presented in Ref.~22.
However, the quark
propagator now has two poles, $i k_0=E_i$.
This is similar to the case of the
Wilson action with $r_\tau < 1$. The two poles are determined from
\bqa
\cosh E_1 &=&\frac{-P - \sqrt{(P^2-QR)}}{Q},\\
{\rm sgn}(Q) \cosh E_2 &=& \frac{-P +\sqrt{(P^2-QR)}}{Q},
\eqa
with additional functions
\beq
P({\vec k})={\cal K}_1 {\cal K}_2 + \kappa_1 \kappa_2,~~~~
Q({\vec k})={\cal K}_2^2 +\kappa_2^2-\delta^2,~~~~
R({\vec k})={\cal K}_1^2 +\kappa_1^2+\delta^2.
\label{PRT}
\eeq
In the limit $\Nt \rightarrow \infty$ the
quark propagator is then given by
\begin{eqnarray}
\hspace*{-0.8cm}S_{\infty}(\tau,{\vec k})\hspace*{-0.2cm} &=&\hspace*{-0.2cm}
\frac{1}{2 \sqrt{P^2-QR} \,\sinh E_1} \\
\hspace*{-0.8cm}
&&\hspace*{-0.2cm}\bigl[ (\kappa_1-i {\cal K}_1) +
(\kappa_2-i {\cal K}_2) \cosh E_1 +\gamma_4 \, \delta \, {\rm sgn}(\tau) \sinh E_1
\bigr] e^{-E_1 \tau} \nonumber \\
\hspace*{-0.8cm}&-&\hspace*{-0.2cm}
\frac{{(-1)}^{\tau \theta(-Q)}}{ 2 \sqrt{P^2-QR} \,\sinh E_2} \nonumber \\
\hspace*{-0.8cm}&&\hspace*{-0.2cm}
\bigl[(\kappa_1-i {\cal K}_1) \,{\rm sgn}(Q)+
(\kappa_2-i {\cal K}_2) \cosh E_2+\gamma_4 \, \delta \,{\rm sgn}(\tau) \sinh E_2
\bigr] e^{-E_2 \tau} . \nonumber
\label{perfect_qp}
\end{eqnarray}
The first term describes the propagation of a physical state
with the energy (dispersion relation) $E_1(\vec k)$ while the
second corresponds to an unphysical state, the analog of the time doubler
in the case of the Wilson action with $r_\tau < 1$.
In Fig. \ref{perfect_disp} we show $E_1$ and $E_2$ as function of $\vec k$.

\begin{figure}[h]
  \begin{center}
\hspace{-0.5cm}
    \includegraphics[width=7.5cm]{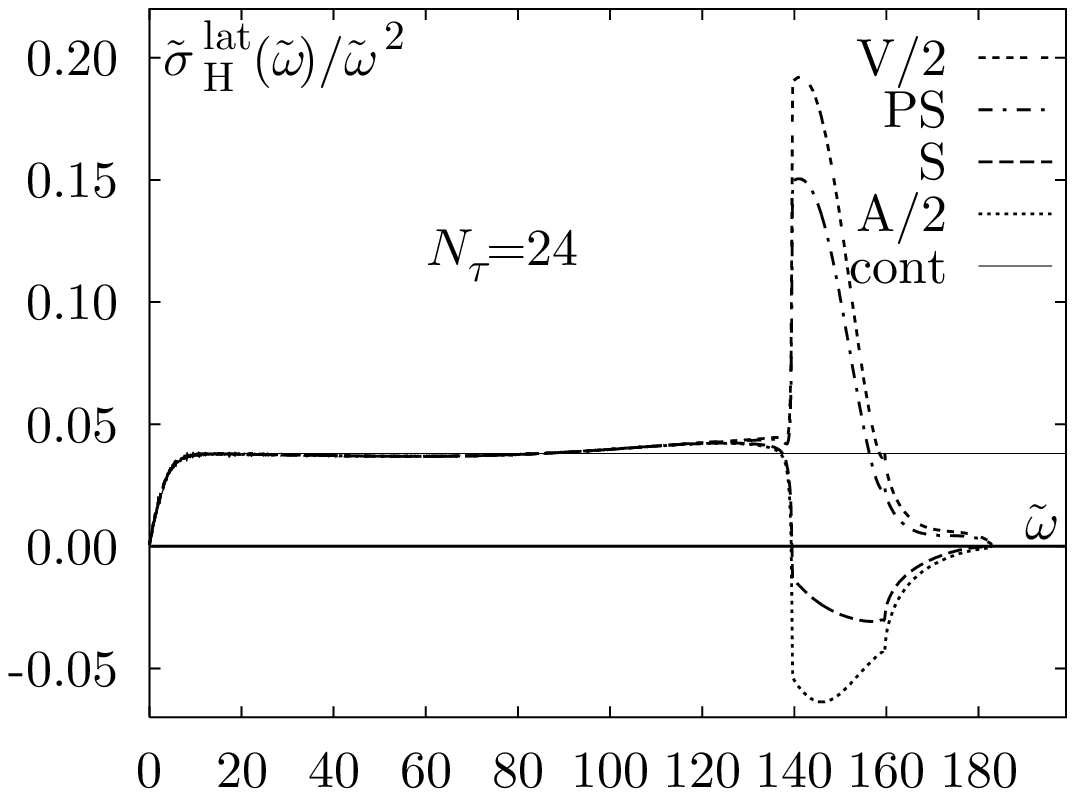}
    \includegraphics[width=7.5cm]{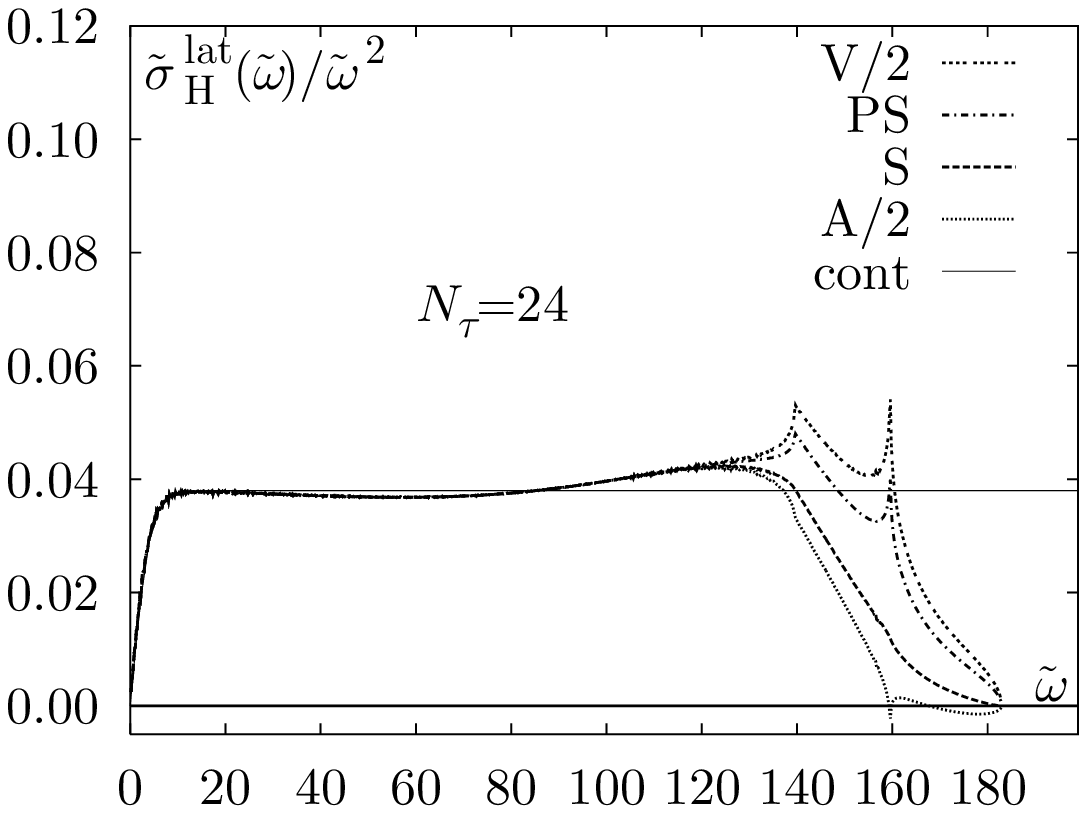}
    \caption{Free spectral functions on isotropic lattices
    calculated with a truncated perfect action for $m=0$.
    The left figure shows the complete result, the right one 
    the contribution from the physical pole ($E_1^p$) alone.
    \label{mfSPF_p}
}
  \end{center}
\end{figure}

As one can see from the figure $E_1(\vec k)$ is very close
to the continuum result for small and moderate momenta
and $E_2$ is much larger than $E_1$.
Only for momenta $|{\vec k}|>2.5$ the gap between $E_2$ and $E_1$ is getting
small. This has important consequences for the meson correlators
which will be discussed below.
Following exactly the same procedure as for the Wilson action
the meson correlators for finite $N_\tau$ can be written as
\bqa
\tilde G_H(\tilde \tau, \tilde p \equiv 0) 
\hspace*{-0.3cm} & = & \hspace*{-0.3cm}
N_c \bigl(\frac{\Nt}{\Ns}\bigr)^3
\sum_{\vec k} \frac{c_H^{lat,p}({\vec k}) \cosh[ 2E_1({\vec k}) \Nt (\tilde \tau-1/2)] +
d^{lat,p}_H({\vec k})}{ (P^2-QR) \cosh^2 (E_1({\vec k}) \Nt /2)} \nn \\
\hspace*{-0.3cm} &   & \hspace*{-0.3cm}
+ \; \Delta G_H^{E_2}(\tilde \tau).
\label{G_H_perfect}
\eqa
The functions $c_H^{lat,p}({\vec k})$ and $d^{lat,p}_H({\vec k})$
are given in Table~\ref{coeff_perfect} in terms of 
$\delta({\vec k})$ and a new function $d^p({\vec k})$,
\beq
d^p \equiv d^p({\vec k})=\frac{({\cal K}_1 + {\cal K}_2 \cosh E_1)^2} 
{ \sinh^2 E_1} \quad .
\label{d1pd2p}
\eeq
The term $\Delta G_H^{E_2}(\tilde \tau)$ in
Eq.~\ref{G_H_perfect} contains contribution from the second pole $E_2$
and an explicit expression for it is given in Appendix B.
As the energies $E_2$ are large it turns out that this part leads to
negligible contributions to the correlation functions except for very
short distances ($\tilde \tau = 0, 1/N_\tau$).
Using the binning-approach discussed in Appendix A we
have calculated numerically the spectral functions in different quantum
number channels. Results for $N_\tau = 24$ are shown in Fig.~\ref{mfSPF_p}.
The good chiral properties and agreement with the continuum
result over a wide range of energies ($\omega a \sim 5$) is self-evident.
The contribution of the physical pole $(E_1^p)$ has also been 
calculated analytically, which can be done in complete analogy to the case of 
the Wilson action. This contribution alone is, as expected, undistinguishable 
from the complete result up to $\omega a \sim 5$,
Fig.~\ref{mfSPF_p}.
\vfill

\begin{table}
\begin{center}
\begin{tabular}{|c|c|c|}
\hline
$H$                &  $c_H^{lat,p}$          &  $d_H^{lat,p}$           \\
\hline
$PS$               &  $\delta^2$            &  $0$                        \\
$S$                &  $- d^p$               &  $d^p-\delta^2$       \\
\hline
 $V_0$             &  $0$                    &  $-\delta^2$            \\
$\displaystyle \sum_{i=1}^3 V_i$     &
                      $3\delta^2-d^p$     &  $d^p$   \\
$\displaystyle \sum_{\mu=0}^3 V_{\mu}$ &
                      $ 3\delta^2-d^p$ &  $ d^p-\delta^2$ \\
\hline
 $A_0$             &  $\delta^2-d^p$      & $d^p$        \\
$\displaystyle \sum_{i=1}^3 A_i$     &
                      $-2 d^p$               & $2 d^p-3 \delta^2$     \\
$\displaystyle \sum_{\mu=0}^3 A_{\mu}$ &
                      $\delta^2-3d^p$     & $3(d^p-\delta^2)$        \\
\hline
\end{tabular}
\end{center}
\caption{The coefficient $c_H^{lat,p}$ and $d_H^{lat,p}$ appearing in 
Eq.~\ref{G_H_perfect}. The functions $d^p$ and $\delta$ are defined in
Eq.~\ref{d1pd2p} and Eq.~\ref{deltadef}, respectively.
}
\label{coeff_perfect}
\end{table}

\section{Conclusions}

We have presented an explicit calculation of mesonic spectral
functions in the infinite temperature limit of lattice QCD.
We have analyzed the cut-off dependence of spectral functions
in different quantum number channels for the Wilson fermion
action on isotropic as well as anisotropic lattices.
We find that the cut-off effects are of similar magnitude in
both cases. The introduction of a Wilson $r$-parameter
being less than unity, in particular the choice $r=1/\xi$ does
not seem to lead to a significant reduction of cut-off effects.

Furthermore, we have analyzed the spectral representation of
mesonic correlation functions using a truncated perfect action.
As expected this does lead to a drastic improvement of 
spectral functions; cut-off effects are shifted to the high
energy regime and chiral symmetry is preserved in the low energy
part of the spectral functions up to energies $\omega a \simeq 5$.

\section*{Acknowledgements}

\noindent
The work has been supported by the DFG under grant FOR 339/2-1
and by the U.S. Department of Energy under Contract DE-AC02-98CH10886.
\vfill\eject

\appendix
\section{Appendix}

\setcounter{table}{0}

We derive here the explicit form of the representation of the free mesonic
lattice spectral functions given in Eq.~\ref{latspec}. In particular,
we give closed analytic expression for the integrals $I_i(y,\xi)$ appearing
in these equations for the case $r=1$ and discuss a representation in
terms of finite Riemann sums which is more convenient for dealing with
the case $r<1$ or spectral functions resulting from more complicated
actions like the truncated perfect action analyzed in Section 5.

\vspace{0.5cm}
\noindent
\underline{$r=1$}: \\[2mm]

We start from Eq.~\ref{G1G2} by performing the obvious variable
transformation ${\vec k} \rightarrow {\vec x}=(\sin^2(k_1/2),\sin^2(k_2/2),
\sin^2(k_3/2))$.
We then define
\begin{equation}
\alpha = {{\cal K}^2 + {\cal M}^2 \over 4 (1 + {\cal M} )}
\label{trans2}
\end{equation}
and rewrite Eq.~\ref{energy} as $E = 2 \ln (\sqrt{\alpha} + \sqrt{1+\alpha} )$.
Using these expressions we can perform a second variable transformation,
$ {\vec x} \rightarrow (\tilde{\omega}, z_2,z_3)=(2N_\tau E, x_2/\alpha,
x_3/\alpha)$.
This leads to the representation of the correlation functions 
$\tilde G_i$ (Eq.~\ref{integraleq})
where the integrals $I_i(y,\xi)$ 
(Eq.~\ref{latspec}) are given by
\begin{equation}
  I_i(y,\xi) = \int\!\!\!\!\!\int_{\Omega(y)} {\rm d}z_2{\rm d}z_3 \;
  A({\vec z})\;B({\vec z})\;C_i({\vec z}) \quad ,
\end{equation}
with ${\vec z}=(z, z_2, z_3)$ and $z(y)=\sinh^2(y/4)$. We further define
\begin{equation}
x = \frac{\xi^2-(z_2+z_3)(1-2\xi z+\hat m)+\xi
\hat m-\frac{\hat m^2}{4z}-2 z z_2z_3}{1+2z(z_2+z_3)+
\hat m-2\xi z} \quad .
\label{xxx}
\end{equation}
With this the functions appearing in the integrand of $I_i$ can be
written as
\begin{eqnarray}
  A({\vec z}) &=& \xi\;\frac{\xi+2z(x+z_2+z_3)+ \hat m}{1+2z(z_2+z_3)+\hat m-2\xi z}
\quad , \\
  B({\vec z}) &=& \left( { 1+2\xi^{-1}z(x+z_2+z_3)+\xi^{-1}\hat m } \right)^{-2}
\quad , \\
  C_1({\vec z}) &=& \left( {x\;z_2\;z_3\;(1-zx)\;(1-zz_2)\;(1-zz_3)
    } \right)^{-1/2} \quad , \\
  C_2({\vec z}) &=& \frac{4}{\xi^{2}} \frac{x(1-zx)+z_2(1-zz_2)+
z_3(1-zz_3)}{(x\;z_2\;z_3\;(1-zx)\;(1-zz_2)\;(1-zz_3))^{1/2}} \quad .
\end{eqnarray}
The boundary of the two-dimensional integration region, $\Omega (y)$, is
given by
\begin{eqnarray}
  \Omega(y)=\left\{ { z_2,z_3\;|\;0\le xy \le 1;\;\; 0\le z_{2}y \le 1 ;
\;\; 0\le z_{3}y \le 1 } \right\} \quad .
\end{eqnarray}

\vspace{0.5cm}
\noindent
\underline{arbitrary $r$}: \\[2mm]

In general we can determine the spectral function in a given quantum
number channel also directly from the representation of the correlation
functions in Eq.~\ref{corrfkt}. We can divide the interval of
non-vanishing energies, $\tilde{\omega} = 2N_\tau E({\vec k})$, in
$n$ bins. Denoting by $\omega_i$ the central values of these bins
and introducing the bin length,
$\epsilon = (\tilde{\omega}_{max} -\tilde{\omega}_{min})/n$,
we can approximate Eq.~\ref{corrfkt} by

\begin{equation}
\hspace*{-0.4cm}\tilde{G}_H(\tilde{\tau},\tilde{p}\equiv0) \simeq
\sum_{i=1}^n \; \epsilon \; s_i \;
\frac{\cosh(\tilde \omega_i (\tilde{\tau}-1/2))}{\sinh(\tilde \omega_i/2)}
\quad ,
\label{corrfktdisc}
\end{equation}
where $s_i$ receives contributions from all terms in the momentum sum
which lead to energies $2N_\tau E({\vec k})$ in the $i$-th bin,

\begin{eqnarray}
s_i =
N_c \left( {\frac{N_\tau}{\xi N_\sigma}} \right)^3
  \sum \limits_{{\vec k}} &~&\hspace*{-0.8cm}
\Theta(2N_\tau E({\vec k}) - (n-1)\epsilon)\;
\Theta(n \epsilon -  2N_\tau E({\vec k}))\cdot \nonumber \\
&&\hspace*{-0.8cm}\frac {c_H^{lat}({\vec k}) \sinh(N_\tau E({\vec k}))}
{(1+{\cal M})^2 \cosh^2(E({\vec k})N_\tau/2)}
~.
\label{bins}
\end{eqnarray}
In the limit of large spatial volumes ($N_\sigma \rightarrow \infty$)
and small bin sizes ($\epsilon \rightarrow 0$) this gives the spectral
functions in a given quantum number channel. We have used this
approach to calculate spectral functions in the case $r<1$ and
also in the case of the truncated perfect action. Typically we used
$N_\sigma \sim 1000$ and $n=1000$.

\section{Truncated perfect action}

In this Appendix we give explicit expressions for the
auxiliary functions which appear in calculations with the
truncated perfect action, in particular in the quark propagator
given in Eq.~\ref{prop_k}.

The explicit form of the functions $\alpha_i({\vec k})$,
$\beta_i({\vec k})$, $i=1,2,3$ and $\delta({\vec k})$,
$\kappa_1({\vec k})$ and $\kappa_2({\vec k})$
are given below
\begin{eqnarray}
&
\displaystyle
\alpha_1({\vec k})=2 \hs_1 (\rho_1+2 \rho_2 (\hc_2+\hc_3)+4 \rho_3 \hc_2 \hc_3)\\
&
\displaystyle
\alpha_2({\vec k})=2 \hs_2 (\rho_1+2 \rho_2 (\hc_1+\hc_3)+4 \rho_3 \hc_1 \hc_3)\\
&
\displaystyle
\alpha_3({\vec k})=2 \hs_3 (\rho_1+2 \rho_2 (\hc_1+\hc_2)+4 \rho_3 \hc_1 \hc_2)\\
[5mm]
&
\displaystyle
\beta_1({\vec k})=4 \hs_1 (\rho_2+2 \rho_3 (\hc_2+\hc_3) + 4 \rho_4 \hc_2 \hc_3)\\
&
\displaystyle
\beta_2({\vec k})=4 \hs_2 (\rho_2+2 \rho_3 (\hc_1+\hc_3) + 4 \rho_4 \hc_1 \hc_3)\\
&
\displaystyle
\beta_3({\vec k})=4 \hs_3 (\rho_2+2 \rho_3 (\hc_1+\hc_2) + 4 \rho_4 \hc_1 \hc_2)
\end{eqnarray}
\beq
\displaystyle
\delta({\vec k})=2 \rho_1 + 4 \rho_2 (\hc_1+\hc_2+\hc_3)+8 \rho_3 (\hc_1 \hc_2+\hc_2 \hc_3+
\hc_1 \hc_3)+16 \rho_4 \hc_1 \hc_2 \hc_3 \label{deltadef}
\eeq
\vspace*{-10mm}
\begin{eqnarray}
&&\hspace{-13mm}\kappa_1({\vec k})=\lambda_0 + 2 \lambda_1 (\hc_1+\hc_2+\hc_3)+4 \lambda_2 (\hc_1 \hc_2+\hc_2 \hc_3+
\hc_1 \hc_3)+8 \lambda_3 \hc_1 \hc_2 \hc_3 \quad\\
&&\hspace{-13mm}\kappa_2({\vec k})=2 \lambda_1+4 \lambda_2 (\hc_1+\hc_2+\hc_3)+8 \lambda_3 (\hc_1 \hc_2+\hc_2 \hc_3+
\hc_1 \hc_3)+16 \lambda_4 \hc_1 \hc_2 \hc_3, \quad
\end{eqnarray} 
where we introduced the shorthand notation $\hc_i=\cos k_i$ 
and $\hs_i=\sin k_i$.

Furthermore, we give the explicit expression for the term in the
meson correlator $\Delta G_H^{E_2}(\tilde \tau)$ coming from contributions
of 
the second pole $E_2$. It can be written as
\begin{equation}
\Delta G_H^{E_2} (\tilde \tau)=G_{H2}(\tilde \tau)+G_{H12}(\tilde \tau)\quad ,
\end{equation}
where $G_{H2}(\tilde \tau)$ contains contributions from the second 
pole $i k_0=E_2$ only and therefore can be written down in close analogy 
with the term containing only the contribution from the first pole $E_1$,
\begin{equation}
\hspace{-0.7cm}\tilde G_{H2}(\tilde \tau, \tilde p \equiv 0)
=N_c \bigl(\frac{\Nt}{\Ns}\bigr)^3
\sum_{\vec k} \frac{c_H^{lat,p2}({\vec k}) \cosh[ 2E_2({\vec k}) \Nt 
(\tilde \tau-1/2)] +
d^{lat,p2}_H({\vec k})}{ (P^2-QR) \cosh^2 (E_2({\vec k}) \Nt /2)}~.
\end{equation}
The functions $P,~Q,~R$ have been defined in Eq.~\ref{PRT};
$d_H^{lat,p2}$ and $c_H^{lat,p2}$ have exactly the same 
structure as $d_H^{lat,p}$ and $c_H^{lat,p}$ listed in 
Table~\ref{coeff_perfect} with $d^p$ being replaced by 
\beq
d^{p2} \equiv d^{p2}({\vec k})=\frac{({\cal K}_1 + 
{\cal K}_2 \; {\rm sgn}(Q) \cosh E_2)^2}
{ \sinh^2 E_2} \quad ,\nonumber \\
\label{d1p2d2p2}
\eeq
The second term $G_{H12}(\tilde \tau)$ contains contributions from 
both the
first and second pole and can be written
as

\begin{eqnarray}
&
\displaystyle
\hspace{-0.5cm}G_{H12}(\tilde \tau)=N_c 
\biggl(\frac{\Nt}{\Ns}\biggr)^3 \sum_{{\vec k}}
\frac{(-1)^{\theta(-Q) \tau+1}}{ (P^2-QR) \cosh(E_1({\vec k}) \Nt/2) 
\cosh(E_2({\vec k}) \Nt/2)}
\nonumber\\
&
\displaystyle
\biggl[ g_H^{lat,p} ({\vec k}) \bigl\{\cosh [E_s({\vec k}) \Nt (\tilde \tau-1/2)]-
\cosh [E_d({\vec k}) \Nt (\tilde \tau-1/2)] \bigr\}
\nonumber\\
&
\displaystyle
\;\;\;\; + \;\;\delta^2({\vec k}) \bigl\{\cosh [E_s({\vec k}) \Nt (\tilde \tau-1/2)]+
\cosh [E_d({\vec k}) \Nt (\tilde \tau-1/2)] \bigr\} \biggr],
\label{GH12}
\end{eqnarray}
where $E_s=E_1+E_2$, $E_d=E_1-E_2$. An additional function 
$g_H^{lat,p}({\vec k})$ has been intro\-duced which depends on 
\begin{equation}
\hspace{-0.5cm}d_1^{p3}=
\frac{{\cal K}_1^2 \,{\rm sgn}(Q)+{\cal K}_2^2 \cosh E_1 \cosh E_2+
{\cal K}_1 {\cal K}_2 ( \cosh E_2+ {\rm sgn}(Q) 
\cosh E_1)}{\sinh E_1 \sinh E_2}\,,
\label{d1p2}
\end{equation}

\begin{equation}
\hspace{-0.5cm}d_2^{p3}=
\frac{\kappa_1^2 \,{\rm sgn}(Q)+\kappa_2^2 \cosh E_1 \cosh E_2+
\kappa_1 \kappa_2 ( \cosh E_2+ {\rm sgn}(Q) 
\cosh E_1)}{\sinh E_1 \sinh E_2}~.
\label{d2p2}
\end{equation}
The explicit form of $g_H^{lat,p}$ for different quantum numbers
is given in Table \ref{tabB}.

\begin{table}[t]
\begin{center}
\begin{tabular}{|c|c|}
\hline
$H$                &     $g_H^{lat,p}$             \\
\hline
$PS$               &   $d_1^{p3}+d_2^{p3}$         \\
$S$                &   $-d_1^{p3}+d_2^{p3}$        \\
\hline
 $V_0$             &  $d_1^{p3}+d_2^{p3}$          \\
$\displaystyle \sum_{i=1}^3 V_i$     &
$d_1^{p3}+3 d_2^{p3}$    \\
$\displaystyle \sum_{\mu=0}^3 V_{\mu}$ &
$2 d_1^{p3}+4 d_2^{p3}$  \\
\hline
 $A_0$             &  $d_1^{p3}-d_2^{p3}$          \\
$\displaystyle \sum_{i=1}^3 A_i$     &
$d_1^{p3}-3 d_2^{p3}$    \\
$\displaystyle \sum_{\mu=0}^3 A_{\mu}$ &
$2 d_1^{p3}-4 d_2^{p3}$  \\
\hline
\end{tabular}
\end{center}
\vspace*{-3mm}
\caption{The explicit form of the functions $g_H^{lat,p}$ appearing
in Eq.~\ref{GH12}. The functions $d_1^{p3}({\vec k})$ and
$d_2^{p3}({\vec k})$ are defined in Eqs.~\ref{d1p2} and \ref{d2p2}.
}
\label{tabB}
\end{table}


\end{document}